# Crystal Growth and High-Pressure Effects of Bi-Based Superconducting Whiskers

*Ryo Matsumoto[a], Sayaka Yamamoto[b,c], Yoshihiko Takano[b,c], Hiromi Tanaka[d]

[a]International Center for Young Scientists (ICYS),
National Institute for Materials Science, 1-2-1 Sengen, Tsukuba, Ibaraki 305-0047, Japan
[b]International Center for Materials Nanoarchitectonics (MANA),
National Institute for Materials Science, 1-2-1 Sengen, Tsukuba, Ibaraki 305-0047, Japan
[c]University of Tsukuba, 1-1-1 Tennodai, Tsukuba, Ibaraki 305-8577, Japan
[d]National Institute of Technology, Yonago College, 4448 Hikona, Yonago, Tottori 683-8502, Japan

Corresponding author
Ryo Matsumoto: MATSUMOTO.Ryo@nims.go.jp

**Abstract**

Three growth methods were tested for producing high-transition temperature superconducting $Bi_2Sr_2Ca_{n-1}Cu_nO_{2n+4+\delta}$ whiskers, employing different ways to focus a compressive stress and size effect of the precursors. First, thermographic imaging was used to investigate thermal stress from temperature distribution in the precursors during growth annealing. To enhance thermal stress in the precursors, a thermal cycling method and an Ag-paste coating method were proposed and found to significantly accelerate the whisker growth. The use of pulverized precursors also promoted whisker growth, possibly due to contribution from the vapor-liquid-solid growth mechanism. The obtained whiskers revealed the typical composition, diffraction patterns, and superconducting properties of the Bi-2212 phase. The proposed methods were able to stably produce longer whiskers compared to the conventional method. Using the obtained whiskers, electrical transport measurements under high pressure were successfully performed at up to around 50 GPa.

# 1. Introduction

For over 70 years, whisker materials consisting of needle-like single crystals have attracted attention as unwanted spontaneous growth on metal electrodes in electronic devices [1-3]. On the other hand, whisker crystals could provide insights into the intrinsic physics of functional materials because of their nearly perfect crystallinity [4]. The growth mechanism of simple substance whiskers has been studied in the past several decades [5-7], but it has not been understood completely. A key factor for whisker growth in several simple substances is considered to be compressive stress in the precursors [8-10].

Recently, $Bi_2Sr_2Ca_{n-1}Cu_nO_{2n+4+\delta}$ (hereafter referred to as "Bi-based") superconducting whiskers have shown superior electrical transport properties [11-15]. Since the Bi-based superconductors exhibit high transition temperatures ($T_c$) of 40, 80, and 110 K at $n = 1$, 2, and 3, respectively [16,17], they have been considered for wire applications to transport electricity without any energy loss [18-20]. The advantage of the Bi-based superconducting wire is easy fabrication as a multifilamentary tape via the powder-in-tube process because of the adjustable feature of the crystal orientation [21,22]. The critical current density ($J_c$) is a major bottleneck for superconducting materials in wire applications, by placing a limit on the current that could pass through them [23]. The developments of Bi-based superconducting wires with high $J_c$ property have been continued for practical use such a training-quench-free coils [24-26]. The most important strategy for enhancing $J_c$ in superconducting wires is to grow strong pinning centers [27,28]. Therefore, investigating the pinning effects for intragrain $J_c$ is required for understanding the intrinsic properties of superconducting materials. Superconducting whisker crystals are useful for investigating the intragrain $J_c$ because their superior crystalline nature precludes the pinning effect from the crystal defects. Recent studies on Bi-based whiskers revealed a drastically enhanced intragrain $J_c$ up to $2 \times 10^5$ A/cm$^2$ beyond the practical criterion for wire applications by introducing pillar-shaped nanocrystalline domains [29]. Partial substitution of $Mg^{2+}$ ion into $Ca^{2+}$ site improved the $J_c$ anisotropy in Bi-based whiskers under an applied magnetic field [30]. These results are quite useful for enhancing the electrical transport properties of not only Bi-based cuprates but also superconducting materials in general. Moreover, the perfect crystallinity of the whisker crystal makes it useful for studying the basic physics of transport properties, such as the superconducting transition temperature and thermoelectric performance [31]. It is also advantageous that the crystal orientation can be determined easily in the whisker due to the specific aspect ratio.

Conventionally, Bi-based whiskers can be obtained using the glassy quenched platelet (GQP) method [32-36]. The proposed mechanism of whisker growth is "microcrucible" based on bottom-end type growth [37]. In the microcrucible mechanism, there are 3 stages of whisker growth. First, the local melts are formed on the surface of GQP during the growth annealing. Second, the nucleation occurs in the macrocrucible. Finally, the whisker can be grown from nuclear. The most growth methods using catalytic additive ($Al_2O_3$ [36], $TeO_2$ [38], $Bi_2O_3$ [39], $Sb_2O_3$ [40], SnO [41], Ga [42], Cu [43], and so on [37]) are within the "microcrucible" mechanism. Recently, compressive stress in the precursor was reported as a hidden factor contributing to the growth of Bi-based whiskers [44]. The formation of "microcrucible" and/or nucleation is considered to be accelerated by the stress. Also, the literature reported a promotion of whisker growth by using crashed precursors [44]. In this study, we demonstrate three methods to promote whisker growth by focusing a compressive stress and size effect of the precursor. Moreover, we investigated high-pressure effects on the electrical transport properties of the obtained Bi-based whiskers.

## 2. Results and discussion
### 2.1 Thermal cycling method

Three different growth methods were applied to the obtained precursors. In the first method (thermal cycling), the precursor was cut to approximately 4 mm × 6 mm × 0.7 mm and placed in an Al$_2$O$_3$ boat as shown in Fig. 1 (a). The boat was placed in a tube furnace heated to 885°C and annealed for 12 hours in an oxygen flow of 120 ml/min. Afterward, the sample was removed from the furnace and cooled in the air for 1 hour to room temperature. Then the sample was annealed again in the tube furnace for 12 hours at the same temperature. This thermal cycling was repeated 5 times, with a total annealing time of 60 hours.

Before the thermal cycling test, temperature distribution during annealing was investigated using a thermographic camera. The Al$_2$O$_3$ boat was first cut cross-sectionally for revealing the heated precursor as shown in Fig. 2 (a). The sample and the boat were then moved into the tube furnace and heated to 885°C. Figure 2 (b) is a thermographic image of the sample during heating. Due to the large heat capacity of the Al$_2$O$_3$ boat, a large temperature difference existed between the edge (400°C) and center (500°C) of the precursor. In this situation, the thermal stress at the center part should be higher than that at the edge part because the thermal stress is proportional to the change in temperature ($\Delta T$) during heating [49]. The gradient of thermal stress induces a deformation of the precursor at the initial stage of growth annealing. According to the previous report [44], the compressive stress induced by the deformation in GQP helps the formation of nucleation of the whisker. The thermographic image first reveals the mechanism of natural deformation of the GQP in the growth process of the whisker. However, the temperature difference disappeared after achieving the annealing temperature of 885°C.

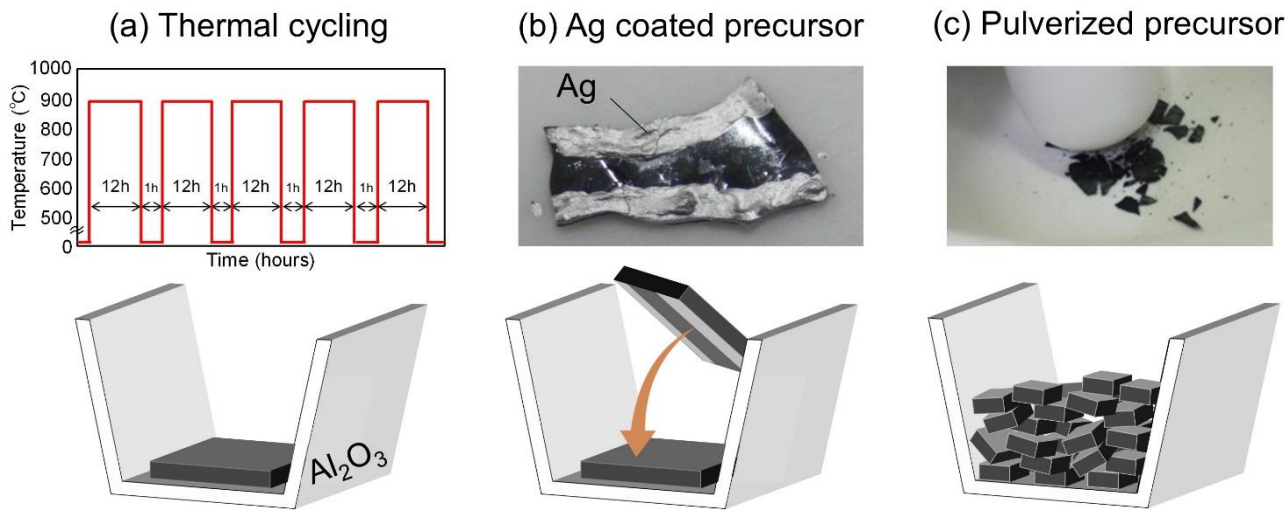

**Figure 1.** Schematics of the (a) thermal cycling method, (b) Ag coating method, and (c) pulverizing method.

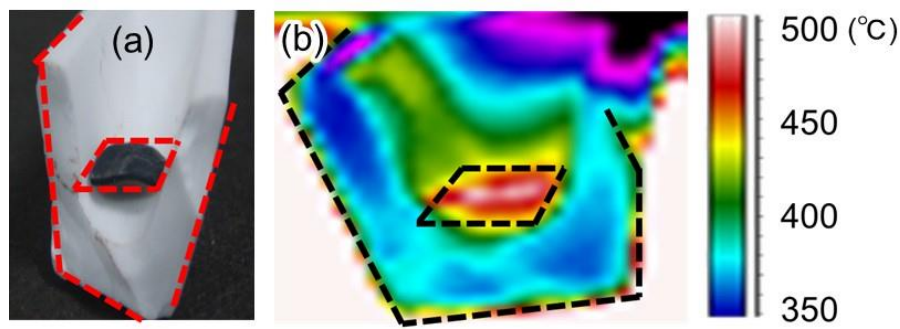

**Figure 2. (a) Optical image of the Al₂O₃ boat with the precursor after a cross-sectional cut. (b) Thermographic image during the heating process.**

The thermal cycling sequence in Fig. 1 (a) was designed to repeatedly induce temperature difference in the precursor. Figure 3 shows optical images of the precursor after each annealing step. The whiskers were partially grown from the precursor after the first cycle, and their number gradually increased after subsequent annealing. After 5 cycles and a total annealing time of 60 hours, there was remarkable whisker growth from the whole region. The shape of the precursor was also drastically deformed after thermal cycling due to the repeatedly induced thermal stress. According to a previous report, compressive stress in the precursors helps form the nuclei for whisker growth [44]. Therefore, the thermal cycling treatment is effective for an increase of the number of whiskers because of an intermittent formation of the nucleation, rather than the increase of whisker length. Since this approach can provide a large number of whiskers from one precursor, high-efficiency fabrication of whisker application is also expected.

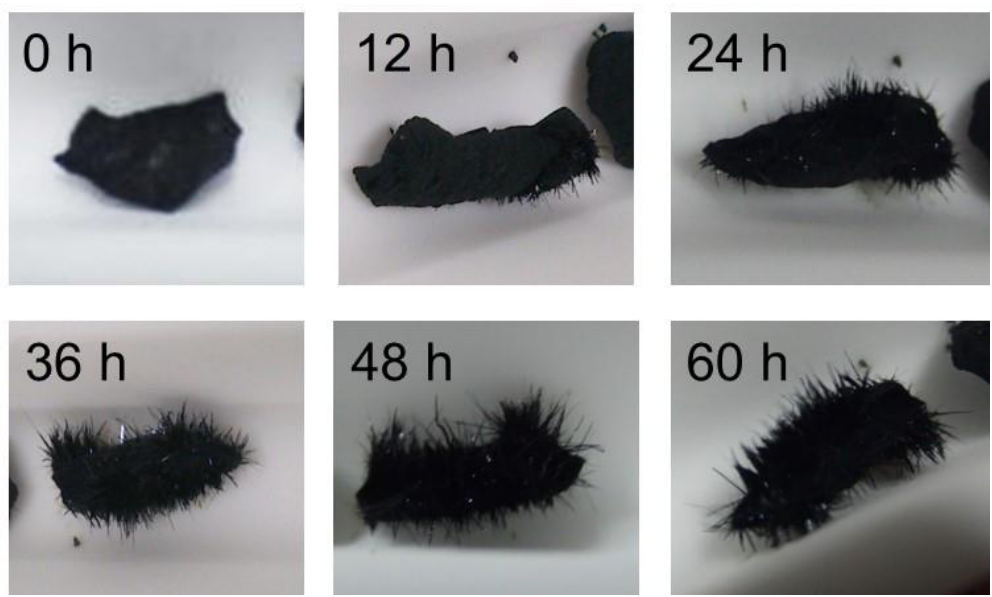

**Figure 3. Optical images of the precursor after each annealing step in thermal cycling.**

**2.2 Ag paste coating method**

The setup used in the second method (Ag paste coating) is depicted in Fig. 2 (b). The precursor without Al₂O₃ catalyst was cut into small pieces, and Ag paste was coated on one side as shown in the inset of Fig. 1 (b). The coated precursor in the Al₂O₃ boat was annealed in a tube furnace at 885°C for 25–100 hours in an oxygen flow of 120 ml/min. The coated Ag paste on one side of the precursor induces thermal stress during the whisker growth process. The volume of Ag paste was reported to shrink by 20% at high temperatures due to evaporation of the solvent [50,51]. Figure 4 displays the optical and SEM images for the (a) as-prepared precursor and (b) Ag paste-coated precursor after annealing for whisker growth. The as-prepared precursor exhibited no remarkable deformation or whisker growth, whereas the paste-coated one was strongly bent and produced whiskers. The whiskers grow from the uncoated surface of the precursor. Figure 4 (c) plots the Ag paste-coated area against the curvature of the precursor after annealing (estimated by circle fitting of the SEM image). The linear relationship between the curvature and the coated area is observed. Although the control of the amount of stress was difficult in the previous report [44], the Ag paste coating exhibits better controllability of the stress.

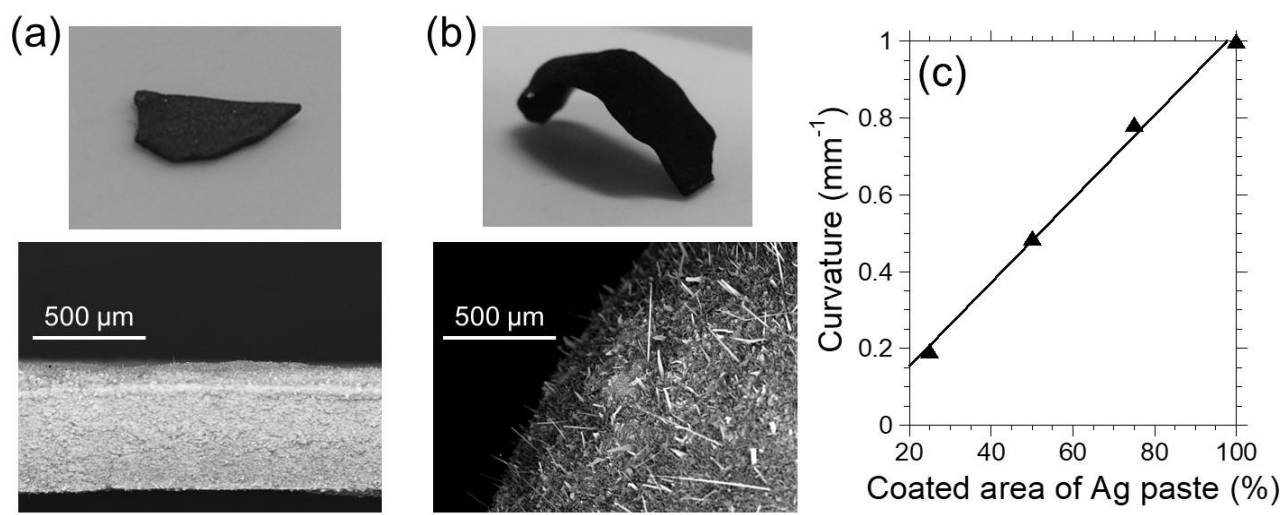

**Figure 4. Optical and SEM images after annealing (a) the as-prepared precursor and (b) the Ag paste-coated precursor. (c) Effect of the Ag paste-coated area on the curvature of annealed precursor.**

Figure 5 (a) plots the relationship between the curvature of annealed precursors (with and without coating with Ag paste) and the maximum length of the grown whisker, to confirm the reproducible effect of thermal stress on whisker growth. When a larger area was coated with Ag paste, the curvature of the precursors tended to increase, and longer whiskers were produced. This trend indicates that the coated Ag paste induces thermal stress in the precursor during the annealing process, which in turn contributes to whisker growth. The growth rates of whiskers were compared for the as-prepared and Ag paste-coated precursors in Fig. 5 (b). By enhancing thermal stress in the precursor, the whisker growth rate drastically increased. However, the growth rate tended to saturate despite the Ag paste coating, because thermal stress only appeared during the initial stage of growth annealing. As confirmed in the thermographic observation during the annealing, the deformation of GQP due to the distribution of thermal stress occurred at the change in temperature ($dT/dt \neq 0$). This fact also suggests that the stress mainly contributes to whisker growth by promoting and/or nucleation. On the other hand, the whisker growth itself occurs at constant temperature ($dT/dt \neq 0$). Meanwhile, continuous whisker growth may require other factors such as catalytic additives. We also note that the whiskers obtained from Ag paste-coated precursors contained no Ag element according to the EDX analysis.

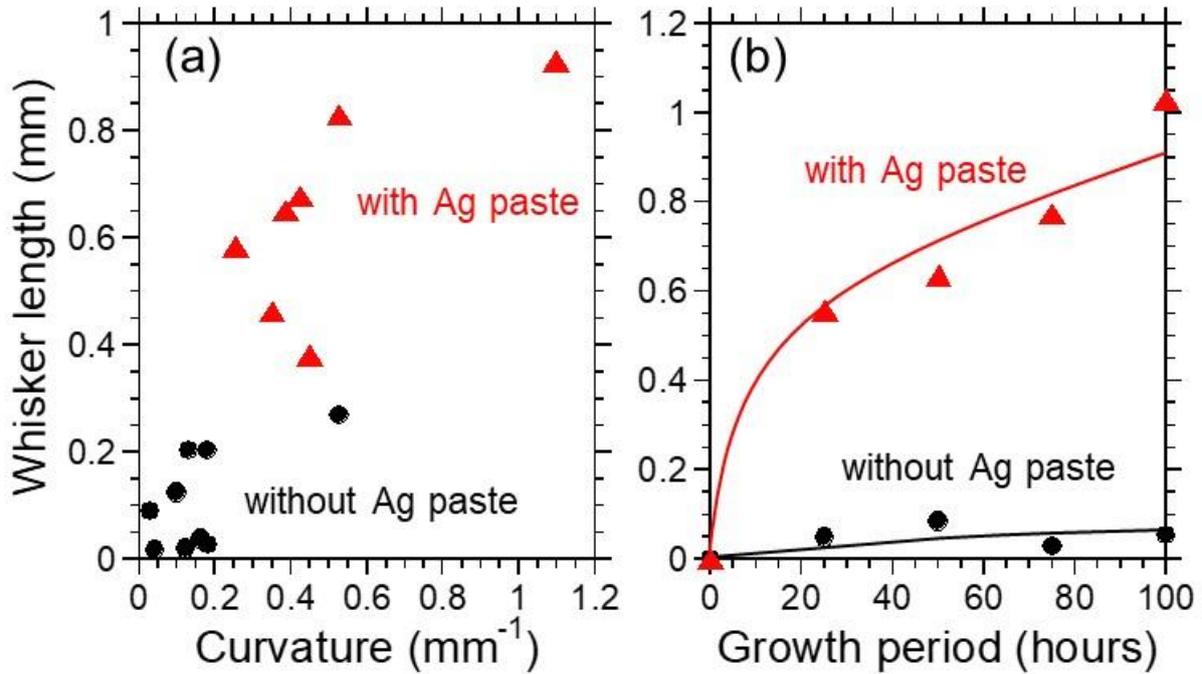

**Figure 5.** (a) Relationship between the curvature of precursor and the length of grown whisker when using the as-prepared and Ag paste-coated precursors. (b) Growth rate of whiskers from both precursors.

**2.3 Pulverizing method**

Figure 1 (c) schematically illustrates the sample setup for the third method (pulverizing). Precursors obtained via the GQP process were ground into rough tips, and the grain size was screened by passing through sieves with 0.18, 0.5, 1.0, 1.4, 1.7, and 2 mm mesh. The screened sample grains were spread on the $Al_2O_3$ boat and annealed in a tube furnace using the same conditions as for the Ag paste coating method. Catalytic $Al_2O_3$ powder was incorporated in the precursors to promote whisker growth. Figure 6 (a) is an optical microscope image of the whiskers obtained from the pulverized precursors, showing many whiskers of sufficient length. The inside of the $Al_2O_3$ boat became partially yellow, indicating vaporization of the starting materials during crystal growth. The SEM image in Fig. 6 (b) reveals that some obtained whiskers had a spiral shape, in contrast to the straight, needle-like crystals obtained the other method. Generally, the Bi-based whisker is grown from the bottom surface, namely the partially molten part of the precursor [52,53]. However, the growth of coiled whisker in some materials, for example, $Si_3N_4$ [54], is attributed to a vapor-liquid-solid (VLS) process. Some studies suggested that the VLS mechanism also contributes to the growth of Bi-based whiskers [55-57]. So, the spiral shape of our whiskers obtained from the pulverized precursors may be the coiled whiskers grown in the VLS process. As discussed in other studies, more reliable evidence for VLS growth could be obtained from an Arrhenius relation between the growth temperature and growth rate [57].

Figure 6 (c) is a typical XRD pattern of whiskers obtained from the pulverized precursors. The (00$l$) peaks of the typical Bi-2212 phase were observed, similar to other Bi-based whiskers [44]. EDX analysis showed that the obtained whiskers had a Ca-rich composition of Bi : Sr : Ca : Cu = 2 : 1.5 : 1.5 : 2, and the same tendency was observed in a previous report [38]. According to the XRD and EDX results, whiskers obtained from the pulverized precursors could be considered typical Bi-2212 whiskers.

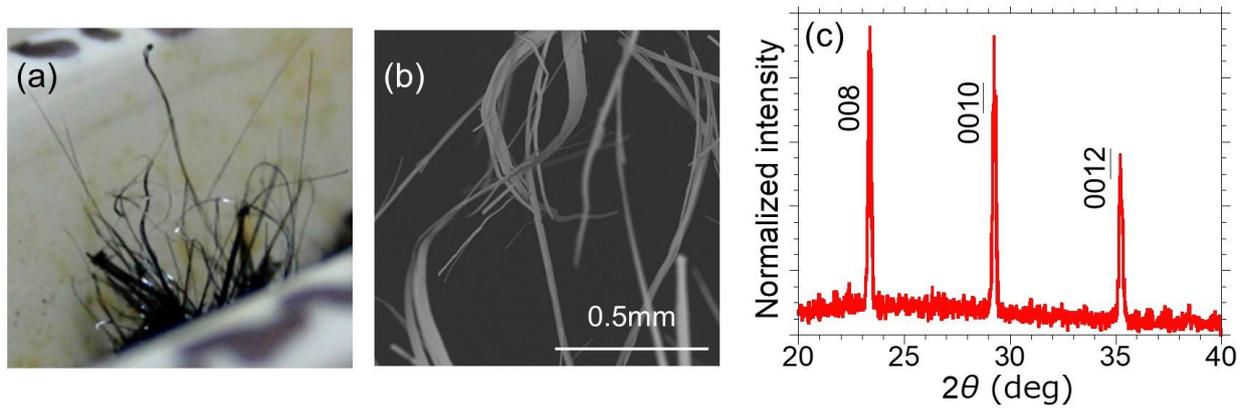

**Figure 6. (a) Optical microscope image and (b) SEM image of whiskers obtained from pulverized precursors. (c) Typical XRD pattern of whiskers obtained from pulverized precursors. The grain size of the precursor is 1 mm × 1 mm × 0.7 mm.**

We grew Bi-based whiskers from both conventional ASGQP precursors and pulverized precursors using various growth period up to 170 hours. The maximum length of the grown whiskers was measured by SEM and compared in Fig. 7 (a), and the growth rates in both methods are presented in Fig. 7 (b). The growth curve for the conventional ASGQP method tends to saturate after 50 hours, and the longest whiskers were ~5 mm. In comparison, our proposed method using pulverized precursors showed no saturation of the growth rate, and the maximum length of the whiskers was much longer (~11 mm). Such a high-efficiency growth of Bi-based whiskers is a remarkable advantage for practical use, compared with the previous methods. Most of the whiskers exhibited a spiral shape, indicating that the VLS process may have contributed to sustained growth without saturation. Taken together, these facts suggest that in the proposed pulverizing method, the thermal stress and VLS-like growth improved the maximum length of the whiskers and the whisker growth rate, respectively. In the future, these factors in whisker growth could be further examined to better understand the growth mechanism and obtain longer Bi-based crystal whiskers.

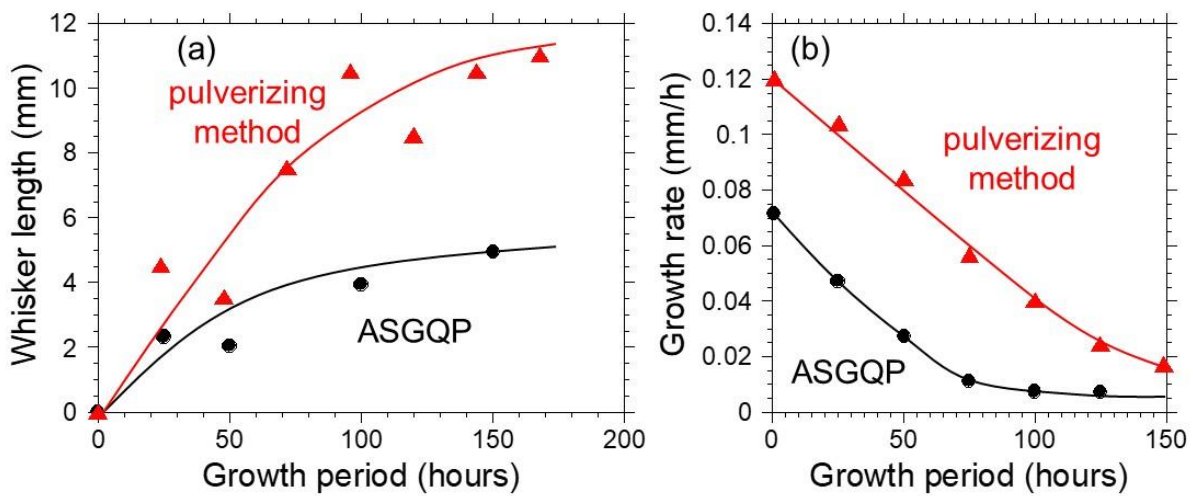

**Figure 7. Comparison of (a) the maximum length and (b) the growth rate of Bi-based whiskers after different growth periods when using conventional ASGQP precursors and pulverized precursors. The grain size of the precursor is 1 mm × 1 mm × 0.7 mm.**

Whisker growth using the pulverizing method was further optimized by testing precursors of different grain sizes. Figure 8 shows the maximum whisker length from precursors sorted by sieves with 0.18, 0.5, 1.0, 1.4, 1.7, and 2 mm mesh. The longest whiskers were grown from the 0.5mm precursors, and the length tended to decrease linearly when further increasing the mesh size. On the other hand, the finest precursors (0.18 mm) provided the shortest whiskers. Here, the whisker growth can be basically explained by the "microcrucible" mechanism [37]. In this growth mechanism, enough space is required at the root of the whisker to form the microcrucible composed of catalytic additive, defects, and so on. When the precursors were cut to very small pieces, there was no space to form the microcrucible, and the whisker growth was impeded.

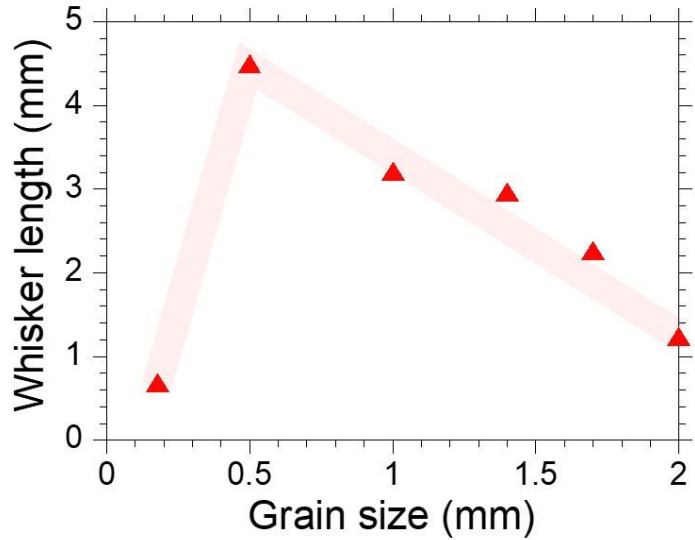

**Figure 8. Maximum whisker length from precursors with different grain sizes.**

Figure 9 (a) shows the typical temperature-dependent electrical resistivity in the obtained whiskers. The normal resistivity from around 110 K to 300 K exhibited the well-known feature of overdoped or optimally doped superconductors [58], a behavior that is consistent with a previous report [59]. The resistivity started to drop sharply below around 110 K, and reached zero at around 80 K. According to the resistivity curve under various magnetic fields up to 7 T (Fig. 9 (b)), both critical temperatures decreased. These results indicate the critical temperature values of 110 K and 80 K, which correspond to the superconducting transition from Bi-2223 phase and Bi-2212 phase, respectively [44]. Figure 9 (c) shows the typical temperature dependence of magnetic moment in the obtained whisker. The magnetic moment mainly dropped at around 80 K, corresponding to the superconductivity transition from Bi-2212 phase. On the other hand, the superconducting signal at 110 K from Bi-2223 phase was quite tiny, according to the inset at an enlarged scale. Such a small volume fraction is a well-known feature of the intergrowth of Bi-2223 phase in the Bi-2212 whiskers [44]. This is consistent with the XRD result, which provided no signal of Bi-2223 phase in the obtained whisker. Despite the small volume fraction of the intergrown Bi-2223 phase, its favored growth direction is the same as that of the Bi-2212 whisker. This caused the sharp drop in the resistivity curve at 110 K.

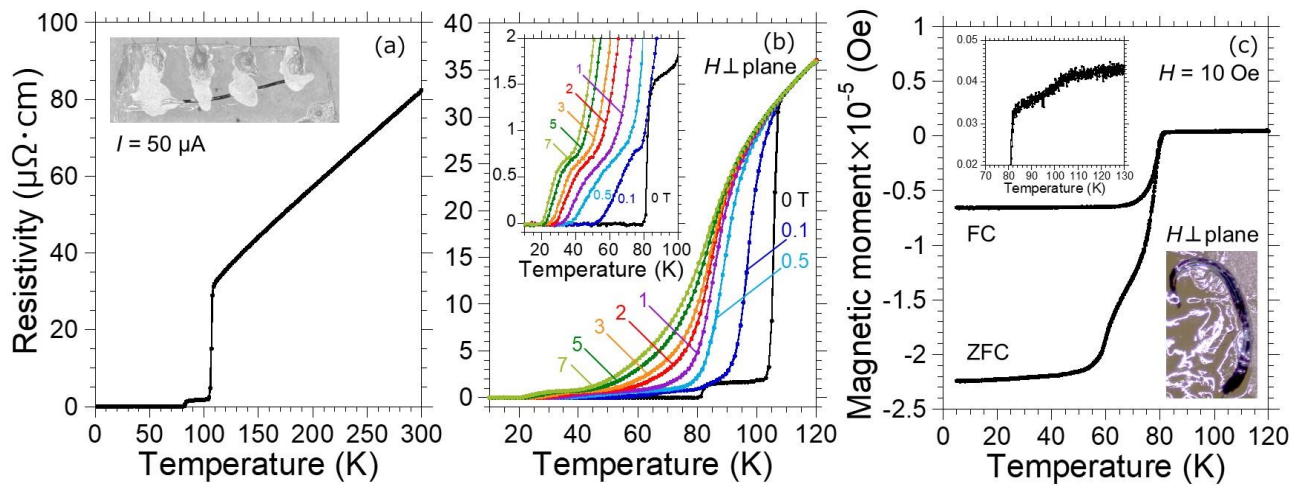

**Figure 9.** (a) Temperature dependence of electrical resistivity in the obtained whisker. (b) Resistivity curves under various magnetic fields. (c) Temperature dependence of magnetic moment in the obtained whisker. Insets of (a) and (c) are photographs of the sample. Magnified regions from (b) and (c) are also shown as insets.

**2.4 High-pressure effects**

The high-pressure effects on superconductivity were investigated for the Bi-based whiskers grown from the pulverizing and pelletizing methods. The temperature dependence of resistance of the whisker was measured under various pressures up to 48.7 GPa. Figure 10 shows the resistance curves from (a) 2.9 to 16.1 GPa and (b) 20.2 to 48.7 GPa. At 2.9 GPa, the resistance started to decrease from 80 K, corresponding to the onset of superconductivity from the Bi-2212 phase. Since the sample used for the high-pressure measurements was tiny, a superconducting transition from the intergrown Bi-2223 phase was not observed. The $T_c$ of Bi-2212 phase gradually decreased when increasing the applied pressure, and then disappeared at 35.2 GPa. Although the measurement continued up to 48.7 GPa, the superconductivity was not recovered. Corresponding to the pressure-driven decrease in $T_c$, the critical current of superconductivity also decreased as shown in the current-voltage curves of Fig. 10 (c).

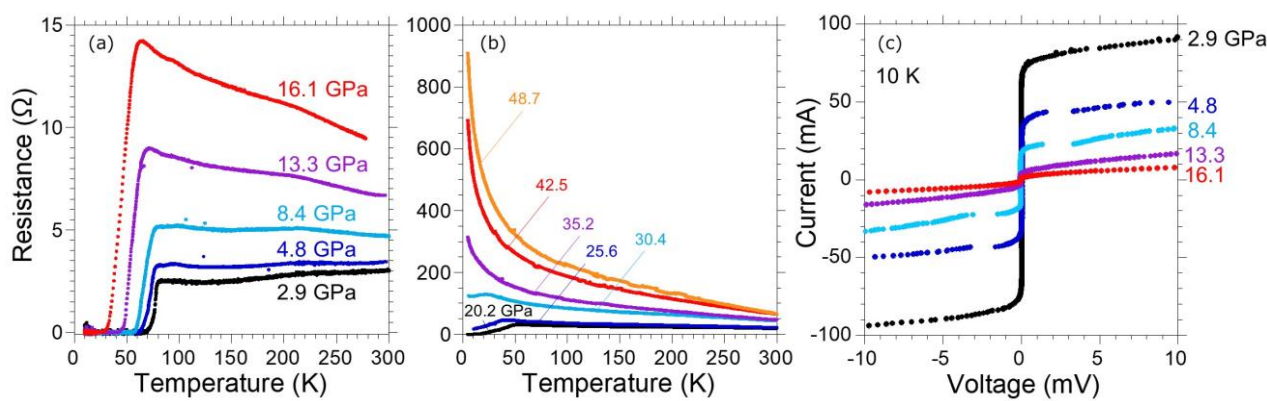

**Figure 10.** Temperature dependence of whisker resistance measured under pressures from (a) 2.9 to 16.1 GPa and (b) 20.2 to 48.7 GPa. (b) Current-voltage curves under various pressures up to 16.1 GPa.

Figure 11 exhibits the effect of applied pressure on $T_c$ in the Bi-based whisker. $T_c$ first increased and then decreased upon increasing the pressure, and this behavior is similar to the trend for the superconducting carrier density in Bi-2212 [60]. According to a recent report, underdoped Bi-2212 and Bi-2223 showed enhanced $T_c$ at high pressures [61]. On the other hand, slightly overdoped Bi-2212 exhibited a decrease of $T_c$ up to around 30 GPa, and then it increased up to around 60 GPa due to an increased density of states at the Fermi level [62]. In our case, the sample showed no sign of enhancement in $T_c$ up to around 50 GPa.

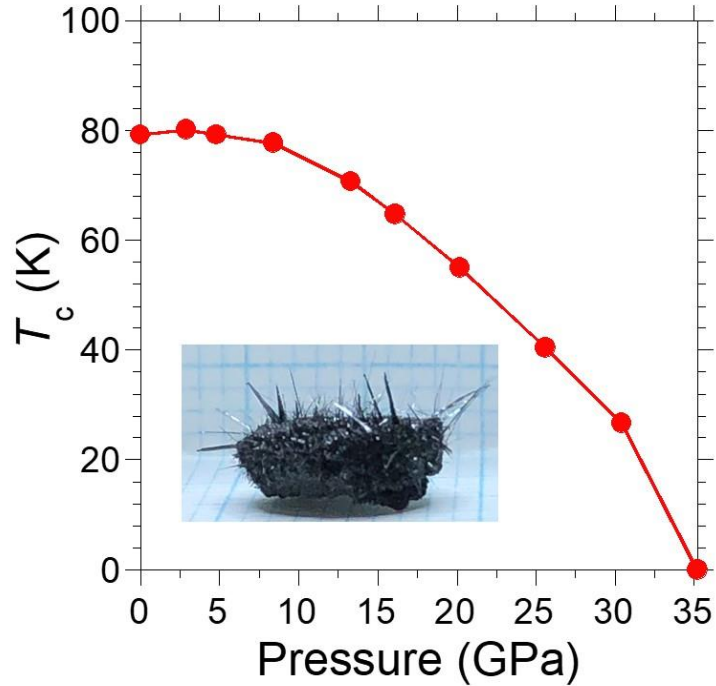

**Figure 11. Dependence of $T_c$ on applied pressure for the Bi-based whisker.**

## 3. Conclusions

Three different methods were applied to grow Bi-based whiskers from the precursor to focus a compressive stress and size effect of the precursors to obtain large high-$T_c$ superconducting whiskers stably. As expected, all three proposed methods promoted whisker growth compared with the conventional method. The method using pulverized precursors was particularly effective, and the grown whiskers exhibited a spiral feature that is known to be caused by a VLS growth mechanism. The superconducting properties of the grown whiskers were measured under high pressures. There was a negative correlation between $T_c$ and pressure below 50 GPa, in contrast to the positive pressure effect on slightly overdoped cuprate. This difference is interesting and requires future investigations on different cuprates, for example, the mercury systems.

## 4. Experimental section

Precursors of the Bi-based whiskers were prepared using the $Al_2O_3$-seeded GQP (ASGQP) method [36], as shown schematically in Fig. 12 (a–d). $Bi_2O_3$ (98.0%), $SrCO_3$ (95.0%), $CaCO_3$ (99.5%), and CuO (95.0%) were mixed in a starting powder with a nominal composition of Bi : Sr : Ca : Cu = 2 : 2 : 2 : 4. The mixed powder (20 g) was moved into an $Al_2O_3$ crucible and heated at 1200°C for 30 min in air. The most of carbonates in the starting powder considered to be evaporated during this process.

The molten mixture was spread on an iron plate covered with Al₂O₃ powder, and then the mixture was rapidly pressed by another iron plate. To confirm the intrinsic effects of the compressive stress, we also prepared the precursor without Al₂O₃ to avoid contribution from the Al₂O₃ catalyst. In that case, a Pt crucible used in the process (a), and Al₂O₃ powder was not scattered on the iron plate in the process (b) in Fig. 12.

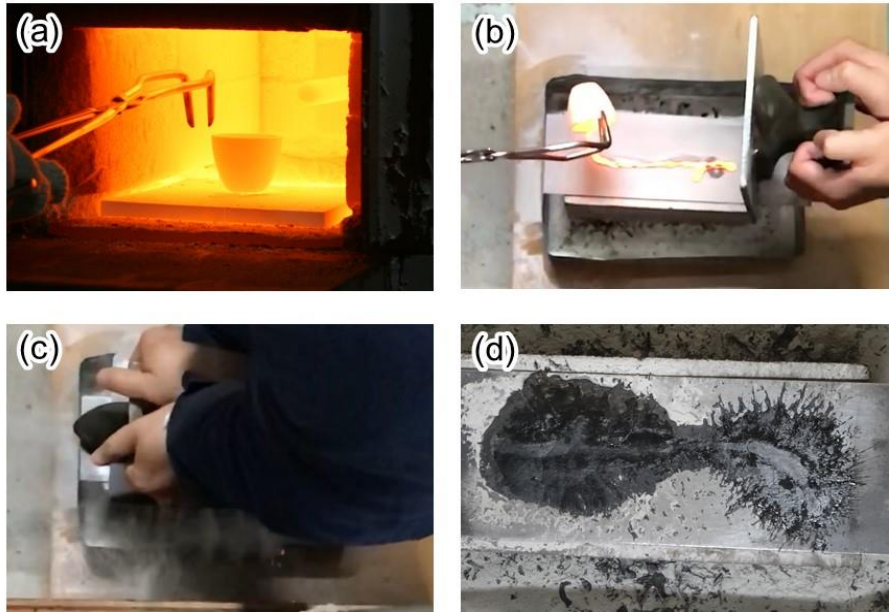

**Figure 12. Steps for preparing the glassy quenched platelet. (a) Starting powder was melted at 1200°C. (b) The molten mixture was spread on an iron plate scattered with Al₂O₃ powder. (c) The mixture was rapidly pressed by another iron plate. (d) Precursor of Bi-based whiskers.**

Temperature distribution during the whisker growth was measured by a thermographic camera (CHINO). The length and composition of the obtained whisker were evaluated by a scanning electron microscope (SEM) equipped with energy-dispersive X-ray spectroscopy (EDX) using TM3000 (Hitachi High-Technologies). The crystal structure was investigated by X-ray diffraction (XRD) patterns measured using Ultima IV (Rigaku) with Cu Kα radiation ($\lambda = 1.5418$ Å). The superconducting properties of the obtained whiskers were evaluated by electrical transport and magnetic susceptibility measurements, carried out using a physical property measurement system and a magnetic property measurement system (Quantum Design), respectively.

Electrical transport in the obtained Bi-based whisker under high pressures was measured in a diamond anvil cell with boron-doped diamond electrodes [45,46]. A whisker obtained using the internal-stress generation method [44] was placed at the center of the bottom diamond anvil, as shown in Fig. 13. Stainless steel and cubic boron nitride were used for the gasket and pressure-transmitting medium, respectively. The metal gasket and boron-doped diamond electrodes were electrically separated by an undoped diamond insulating layer. The sample chamber was compressed by squeezing the other diamond anvil. Pressure in the sample chamber was estimated by the ruby fluorescence method [47] in the low-pressure region and by the diamond-Raman spectroscopy method [48] in the high-pressure region.

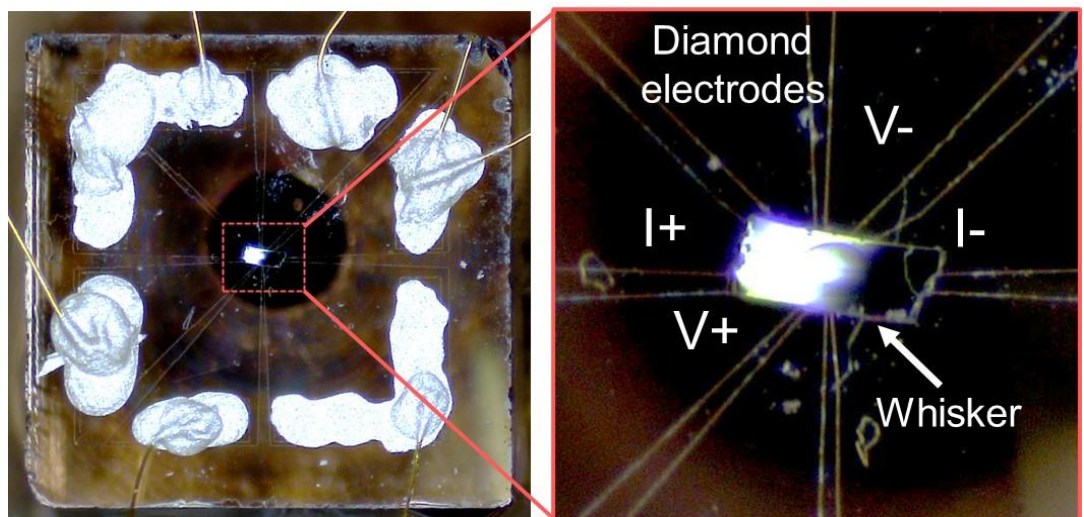

**Figure 13.** Optical images of the prepared diamond anvil with boron-doped diamond electrodes for measuring electrical transport in the Bi-based whisker under high pressures.

**Acknowledgment**

The authors thank Mr. S. Tsunashima, Mr. K. Kamimoto, Ms. K. Maeda, and Ms. T. Nishio for supporting the sample preparation and measurements. This work was partially supported by JSPS KAKENHI (Grant Nos. JP17K06362, JP17J05926, JP19H02177, 20H05644, and 20K22420), JST CREST (Grant No. JPMJCR16Q6), and JST-Mirai Program JPMJMI17A2. The fabrication of diamond electrodes was partially supported by NIMS Nanofabrication Platform in Nanotechnology Platform Project sponsored by the Ministry of Education, Culture, Sports, Science and Technology (MEXT), Japan. The authors would like to acknowledge the ICYS Research Fellowship, NIMS, Japan.